 \documentstyle[twocolumn,prl,aps,epsf]{revtex}     

\begin{document}

\title{Poincar\'e's Recurrence Theorem and 
       the Unitarity of the \mbox{\boldmath $S$}--Matrix}

\author{A. M. Ozorio de Almeida and  R. O. Vallejos}

\address{Centro Brasileiro de Pesquisas F\'{\i}sicas \\
         Rua Xavier Sigaud 150, 22290--180~Rio de Janeiro, Brazil \\}

\date{\today}

\maketitle

\begin{abstract}
A scattering process can be described by suitably closing the system
and considering the first return map from the entrance onto itself.
This scattering map may be singular and discontinuous, but it will be
measure preserving as a consequence of the recurrence theorem applied
to any region of a simpler map.  In the case of a billiard this is the
Birkhoff map.  The semiclassical quantization of the Birkhoff map can
be subdivided into an entrance and a repeller.  The construction of a
scattering operator then follows in exact analogy to the classical
process. Generically, the approximate unitarity of the semiclassical
Birkhoff map is inherited by the $S$--matrix, even for highly 
resonant scattering where direct quantization of the scattering map
breaks down.
\end{abstract}

\draft\pacs{ PACS numbers: 05.45.+b, 03.65.Sq, 03.80.+r }

\narrowtext

For a classical conservative system, whether discrete or continuous in
time, Poincar\'e's celebrated theorem can be reduced to the general
statement that the probability for an orbit to return to any given
region is unity if the motion is bounded. There is no restriction on
the time this recurrence will take, which may vary widely among orbits
starting in different subregions.  In any case, by waiting a
sufficiently long time, the first return of each trajectory defines a
measure preserving map of the region onto itself.

The essential boundedness of the system in no way hinders its
application to classical scattering problems, because we can always
choose the appointed region to coincide with the opening of the
scattering system to the outside world. Since we are only interested in
the first return of the orbits to the chosen region, it makes no
difference that the system is not really closed.  In other words, we
can still apply the theorem if the union of the scatterer and the
opening combine to form a bounded measure preserving map.

As a first example consider the simple scattering system composed of a
circular billiard opening onto a straight tube\cite{Blumel}, as shown
in Fig.~1.  In this case, the closure of the dynamical system can be
reduced to the Birkhoff map (or bounce map) for specular collisions of
the straight trajectories with the circular boundary. The phase space
is defined by the boundary coordinate $s$ (or the angle $\theta$, in
the case of unit radius) and $p_s$, the tangential momentum
(proportional to $\cos\alpha$, where $\alpha$ is the angle of
incidence).

The closed dynamics is very simple in this case:  $p_s$ is constant
(integrable motion) and $\Delta \theta= 2 \alpha$.  Nonetheless, the
scattering map of the orbits returning to the opening is
discontinuous.  Indeed it is composed of an infinite sequence of
diminishing subregions, of which the first few are shown in Fig.~2.
Therefore a maximally simple closed dynamics induces a relatively
complex (resonant) scattering map. It is only in the (non resonant)
limit where the size of the opening approaches the diameter of the
circle that the scattering map is also simple.

Consider now the less obvious example of the specular scattering from
three disks, that has become the paradigm of chaotic classical
scattering\cite{3-disks}.  It may appear that our choice of closing
surface in Fig.~3 amounts to an overkill, since we are not interested
in orbits such as {\bf a} in Fig.~3.  However this integrable motion
described in the previous example does not mix with the scattering
orbits such as {\bf b}, so we can substract it from the phase space of
our system. This is then composed of the Birkhoff coordinates of the
external circuit, restricted to small $p_s$ added to the full Birkhoff
coordinates of the three scattering circles, as shown in Fig.~4.
Evidently, we obtain the useful asymptotic scattering picture by making
the radius of the outer circle arbitrarily large, so that we can
identify the exit direction of an orbit with the point where it
collides with the outer circle. Even so, the useful area for the outer
circle in the phase space of Fig.~4 will be smaller than that of the
three disks combined.

The dynamics for the first return of the closed map is indicated by the
different regions in Fig.~4. This is less trivial than in the previous
example, but it is nowhere singular. The first return map between the
four circles is hyperbolic and discontinuous similarly to the baker's
transformation\cite{baker}, but the full complexity of the motion only
arises through the multiple iterations needed to compose the scattering
map of first returns to the outer circle. This map exhibits a fractal
structure of singularities generated by motion that nearly enters on
the stable manifold of periodic orbits within the
scatterer\cite{fractal}.

In both our simple examples it may still be necessary to relate the map
restricted to the opening to a map purely determined by the mesurement
to be performed. In the case of the circle that opens onto the tube,
one is finally concerned with the orbits entering and leaving the other
end of the tube, whereas in the three disc case we should connect the
enclosing circle to an asymptotically large circle. The resulting map
is known as the Poincar\'e scattering map\cite{Jung}. Our immediate
concern will be the quantization of the (first return) internal map
rather than the details of its outer connections.

An evident conceptual advantage is achieved by understanding the
structure of scattering maps on the basis of multiple iterations of
their closure, even though it may be necessary to reverse this
procedure in experimental situations. Our purpose is now to show that
we can transfer to semiclassical scattering a construction of open and
closed systems corresponding to the one which we have been employing in
the classical theory.

The starting point is to note that we can always define a finite
Hilbert space that will correspond semiclassically to a finite phase
space\cite{Bogomolny}.  Indeed, the dimension of the Hilbert space
corresponding to a two dimensional classical phase space of area $A$
will be $N=A/2\pi\hbar$, where $\hbar$ is Planck's constant. The
prescription given by Miller\cite{Miller} for the approximately unitary
quantum map $U$, is given in the coordinate representation as
\begin{equation}
U(q,q') \simeq \frac{1}{\sqrt{2\pi\hbar}}
\sum_j \left|\frac{\partial^2 \sigma_j}{\partial q \partial q'}
\right|^{1/2}  e^{i \sigma(q,q')/\hbar + i \mu_j} ~,
\label{eqmiller}
\end{equation}
where $\mu_j$ is the Maslov phase and $\sigma_j(q,q')$ stands for the 
generating function of the classical map, given implicitly by
\begin{equation}
p'= \frac{\partial \sigma_j}{ \partial q'} ~,~~
p =-\frac{\partial \sigma_j}{ \partial q } ~,
\end{equation}
the index $j$ indicating that there may be more than one orbit for a
given pair of points $(q',q)$. Of course, we should worry about the
discretization of coordinates, due to the finite dimension ($N$) of the
Hilbert space, but the approximations will hold when $N$ is large.
Furthermore, the semiclassical approximation (\ref{eqmiller}) will
leave out evanescent modes\cite{evanescent} and diffraction effects
({\em e.g.} in the three disks problem), but again these will give
relatively small contributions in the large $N$ limit.

We can check the approximate unitarity of (\ref{eqmiller}) by 
noting that, for each continuous region of area $A_j$ 
\begin{equation}
N_j = \mbox{Tr} \, U_j U_j^\dagger \simeq
      \int \frac{dq \, dq'}{2\pi\hbar}
      \left| \frac{\partial^2 \sigma_j}{\partial q \partial q'} 
     \right|=\frac{A_j}{2\pi\hbar} ~.
\end{equation}
Thus, we are quantizing separately each subregion in a way that
increases border effects for discontinuous maps with many subdivisions.
We have shown that this is typical of scattering maps.  If they are
sufficiently resonant as in our examples, we obtain $N_j
\stackrel{<}{\sim}1$ for many subregions, which are hence beyond the
range of validity of the Miller prescription. The quantum signature of
this resonance problem is the need to account for a large number of
evanescent modes that cannot be related to real classical orbits.

It is important to note that a given scattering experiment may involve
the amplitude of a single element of the $S$--matrix in the appropriate
representation. We thus require detailed local knowledge of this
operator, rather than merely its traces or other coarse grained
information. Experience with the quantized baker's map and other
discontinuous maps analogous to the present scattering situation, shows
that an iteration of the map may have a reasonable semiclassical
approximation for its trace beyond the time when the semiclassical
approximation for the full map has broken down\cite{da Luz}. Moreover,
we are not concerned with the small departure from unitarity of the
Miller construction for a continuous classical map. Instead we seek to
remedy its complete breakdown for highly resonant scattering maps.

The way out is to rely on the construction of the scattering map from
the multiple iterations of the simpler closed map. We thus need the
following result, which may be considered as the quantization of the
recurrence thorem:

Given a finite Hilbert space $H_N$ subdivided into two orthogonal
subspaces 
$ H_{N_0} = P_0 H_N $ and $ H_{N_1} = P_1 H_N $ 
(such that the projection operators $ P_0 + P_1 = 1_N $) 
and given an unitary operator $U_N$ defined on $H_N$, then the operator
\begin{equation}
S_{N_0} = P_0 U_N \left[ 1 - P_1 U_N \right]^{-1} P_0 =
          P_0 U_N \sum_{m=0}^{\infty} \left( P_1 U_N \right )^m P_0
\label{theorem}
\end{equation}
is unitary on $H_{N_0}$. It should be noted that this is a much
stronger property than the obvious ``weak unitarity'' statement
\begin{equation}
\forall \psi_0 \in  H_{N_0}: | \psi_0 |^2 = \sum_{m=0}^{\infty}  
                 | P_0 U_N  \left( P_1 U_N \right )^m  \psi_0 |^2 ~,
\label{weakunitarity}
\end{equation}
which corresponds to the overall conservation of {\em classical}
probability.  To outline the proof of the theorem, we define the
rectangular blocks of the operator $U_N$, namely $U_{ij} = P_{i} U
P_{j}$, and their hermitian conjugates $U_{ij}^\dagger = P_{j}
U^\dagger P_{i}$.  The unitarity of $U_N$ implies
\begin{eqnarray}
 U_{00} U_{00}^\dagger + U_{01} U_{10}^\dagger & = & 1_{N_0}  \\
 U_{00} U_{01}^\dagger + U_{01} U_{11}^\dagger & = & 0        \\
 U_{10} U_{01}^\dagger + U_{11} U_{11}^\dagger & = & 1_{N_1}  ~, 
\end{eqnarray}
where $1_{N_j}$ coincides with the nonsingular block of $P_j$.  In this
notation Eq.~(\ref{theorem}) reads $ S_{N_0}=U_{00} +
U_{01}[1_{N_1}-U_{11}]^{-1}U_{10} $.  Then it is straightforward,
though a little lengthy, to verify that $ S_{N_0} S_{N_0}^\dagger =
S_{N_0}^\dagger S_{N_0} = 1_{N_0} $ is a consequence of Eqs.~(6--8).

Consider now a basis for $H_{N_1}$ in which $P_1U$ is diagonal,
with states $\psi^j_1$, and the corresponding eigenvalues
$\lambda^j_1$. If the elements of $U$ that couple such a state to
the subspace $H_{N_0}$ are labeled $U_{10}^{jk}$, then 
\begin{equation}
|\lambda^j_1|^2 = 1 - \sum_{k} |U_{10}^{jk}|^2~.
\label{lambda}
\end{equation}
It follows that each eigenvalue of $P_1U$ lies inside the unit 
circle, unless the corresponding eigenstate is completely uncoupled
to $H_{N_0}$ (in which case this eigenstate should be substracted
from $H_{N_1}$). The quantum map $P_1U$ is therefore strictly 
dissipative, corresponding to the
classical scatterer that looses orbits at each iteration\cite{smale}. 
This fact
is essential for the convergence of the sum in (\ref{theorem})
of the expansion for $S_{N_0}$.

We can now apply the exact result (\ref{theorem}) to scattering by 
identifying $U_N$ with the approximate semiclassical map
(\ref{eqmiller}) for the closure of the scattering system. The
resulting scattering matrix given by (\ref{theorem}) determines the
on--shell $S$--matrix for fixed energy\cite{Feshbach}.
 
If the semiclassical approximation for $U$ departs from unitarity
by order $\epsilon$, the scattering matrix will err by an order 
$\epsilon[1-U_{11}(\epsilon)]^{-2}$. So there could be a large 
deviation of the $S$--matrix from unitarity if one of the eigenvalues 
of $U_{11}$ were sufficiently close to the unit circle.
However, $\epsilon$ vanishes as $\hbar \rightarrow 0$, whereas the
coupling of the scatterer to the opening is classically strong in 
our examples, so that (\ref{lambda}) guarantees that asymptotically
$[1-U_{11}(\epsilon)]^{-2}$ remains finite. Therefore we can use
the quantum recurrence theorem as a basis for the semiclassical
approximation of the scattering matrix. 

In our examples, the energy dependence of the classical map is trivial
and it can be scaled away, but the area of the phase space grows with
energy, modifying the dimension of the corresponding Hilbert space.
Another way to see this important energy dependence of the quantum
mechanics is through the growth of the actions of the orbits, i.e. the
generating functions $\sigma_j$ in (\ref{eqmiller}). In the case of
smooth potentials, rather than billiards, even the energy dependence is
nontrivial, but these scattering systems can also be treated within our
conceptual framework by introducing quantum Poincar\'e sections in the
manner of Bogomolny\cite{Bogomolny} or Rouvinez and
Smilansky\cite{Smilansky}.

Formally, we could reobtain the standard Miller formula for the
scattering amplitudes by doing the matrix multiplications in the
infinite expansion (\ref{theorem}) by the method of stationary phase.
If we keep the semiclassical representation (\ref{eqmiller}) for the
unitary matrix $U$ and compute
$\left[ 1_{N_1} - U_{11} \right]^{-1}$ 
using ``Fredholm Theory'', we rederive precisely the semiclassical
scattering theory of Georgeot and Prange\cite{Georgeot}. 
Our approach shows that each of the operators employed in the 
semiclassical Fredholm theory originates in the quantization of 
Poincar\'e's recurrence theorem, as well as elucidates 
their approximate unitary or dissipative character.

Finally, our results allow for a minimal semiclassical approximation
for the scattering matrix. The only approximation is the use of Miller's
theory for the global map $U_N$ (rather than for the scattering map, as
in previous treatments), using the classical variables corresponding to
the entrance and exit channels. The problem is then reduced to the
numerical inversion of the matrix $\left[1_{N_1} - U_{11}\right]$, which
has the dimension of the classical phase space of the scatterer divided
by Planck's constant.

\acknowledgements

We would like to thank C. Anteneodo and R. Markarian for useful
suggestions. This work was supported by the Conselho de Desenvolvimento
Cient\'{\i}fico e Tecnol\'ogico (CNPq/Brazil).

\newpage


\begin{figure}[h]
\epsfxsize=8.5cm
\epsfbox[48 158 565 514]{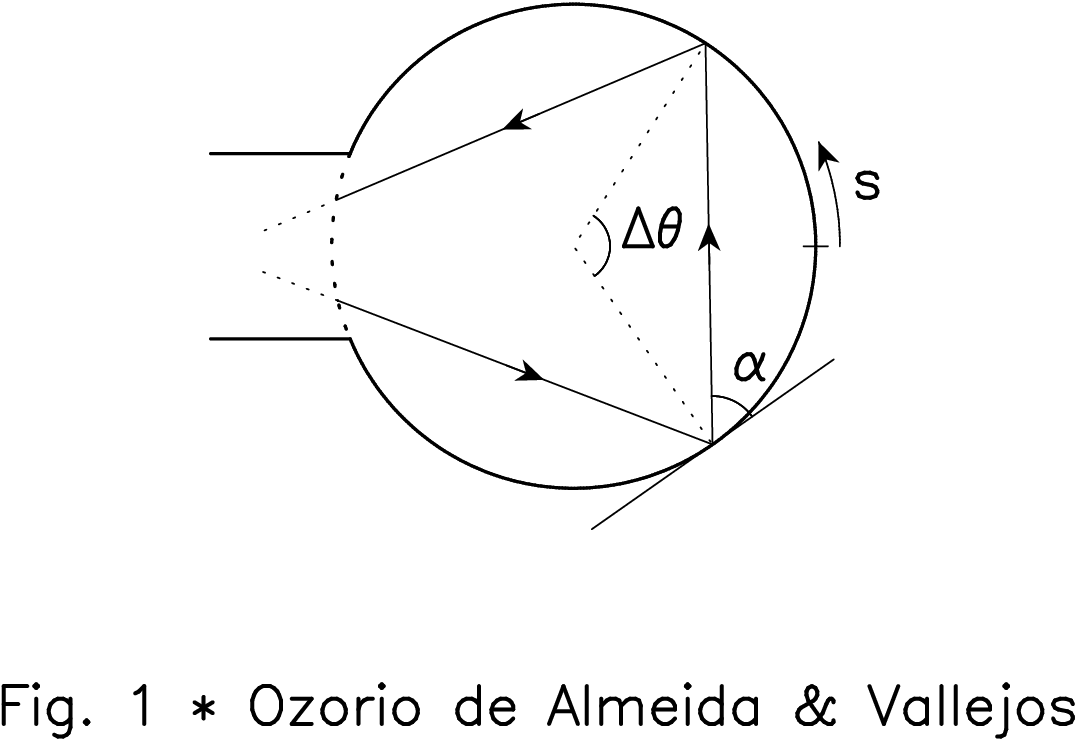}
\vspace*{1.0pc}
\caption{Scattering by a circular cavity.  
A particle propagates from the asymptotic region 
(straight tube) 
up to the entrance of the interaction region 
(dotted arc, approximating a straight segment). 
This defines an
initial condition for
the internal map of specular reflections
against the circular boundary.
Eventually, the particle 
returns to the opening, entering the tube instead of reflecting again.
The coordinates for the internal bounce map are $s$,
the distance along the boundary, and the tangent momentum $p_s$,
proportional to $\cos\alpha$. The rule for specular reflections is
$\Delta\theta=2\alpha$.}
\end{figure}


\begin{figure}[h]
\epsfysize=8.5cm
\epsfbox[48 109 565 593]{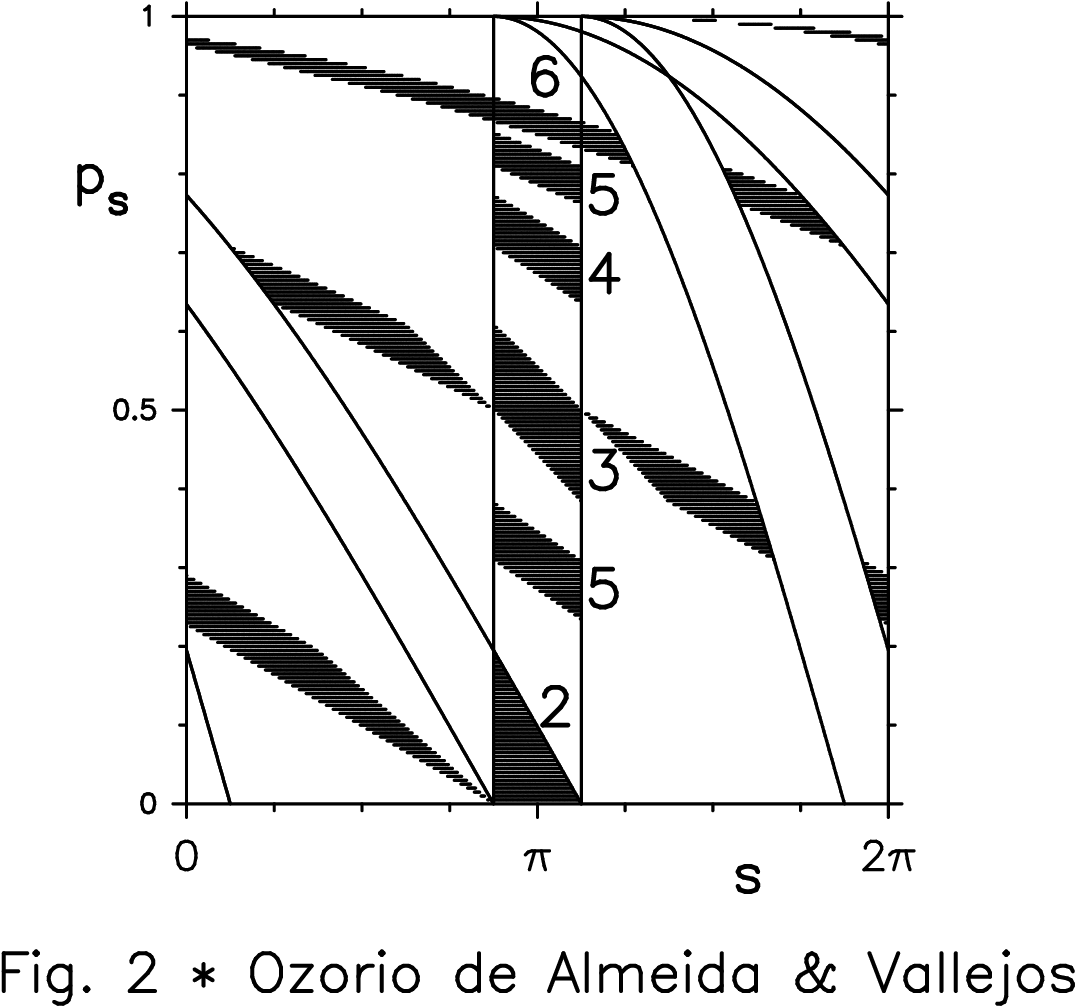}
\vspace*{1.0pc}
\caption{
The first--return map for the circular system of Fig.~1. 
Initial conditions having all possible momenta are launched from 
the entrance of the scattering system.
In the Birkhoff plane $s$--$p_s$, this corresponds to the 
narrow rectangular
region centered at $s=\pi$.
We show the first two images of this region by the full internal 
map (strips bounded by the cosine curves).
Also shown are the first returns
after $2,3,\ldots,6$ bounces with the circular boundary, and the 
phase points that still stay inside the cavity after 6 bounces 
(dark strips ouside the initial rectangle).
We have not plotted those points that return after one iteration as
they do not correspond to the scattering process
(for the system closed by a straight
segment there would be no orbits leaving at the first return).
The picture for $p_s<0$ is obtained by reflection with respect to
the point $(s=\pi,p_s=0)$.} 
\end{figure}


\begin{figure}[h]
\epsfysize=8,5cm
\epsfbox[48 135 565 591]{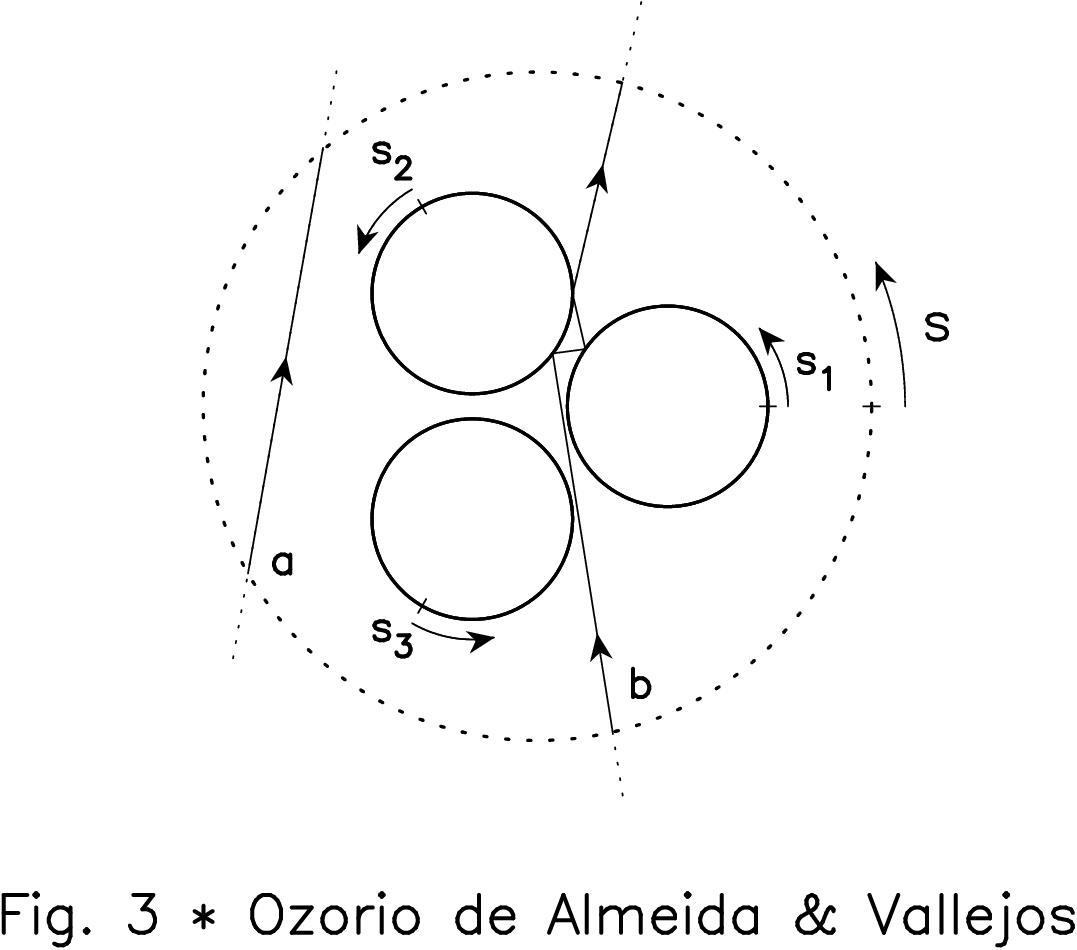}
\vspace*{1.0pc}
\caption{Scattering by three discs. 
An incident particle propagates trivially to the outer circle
that closes the system. 
This defines an initial condition for
the Birkhoff map of the closed problem, described by the full set of
position coordinates $\{s,s_1,s_2,s_3\}$ 
together with the tangent momenta. 
After colliding with the discs a certain number of times, 
the particle returns to the outer circle, {\em i.e.} it escapes.}
\end{figure}

\newpage


\begin{figure}[h]
\epsfxsize=8.5cm
\epsfbox[39 91 573 652]{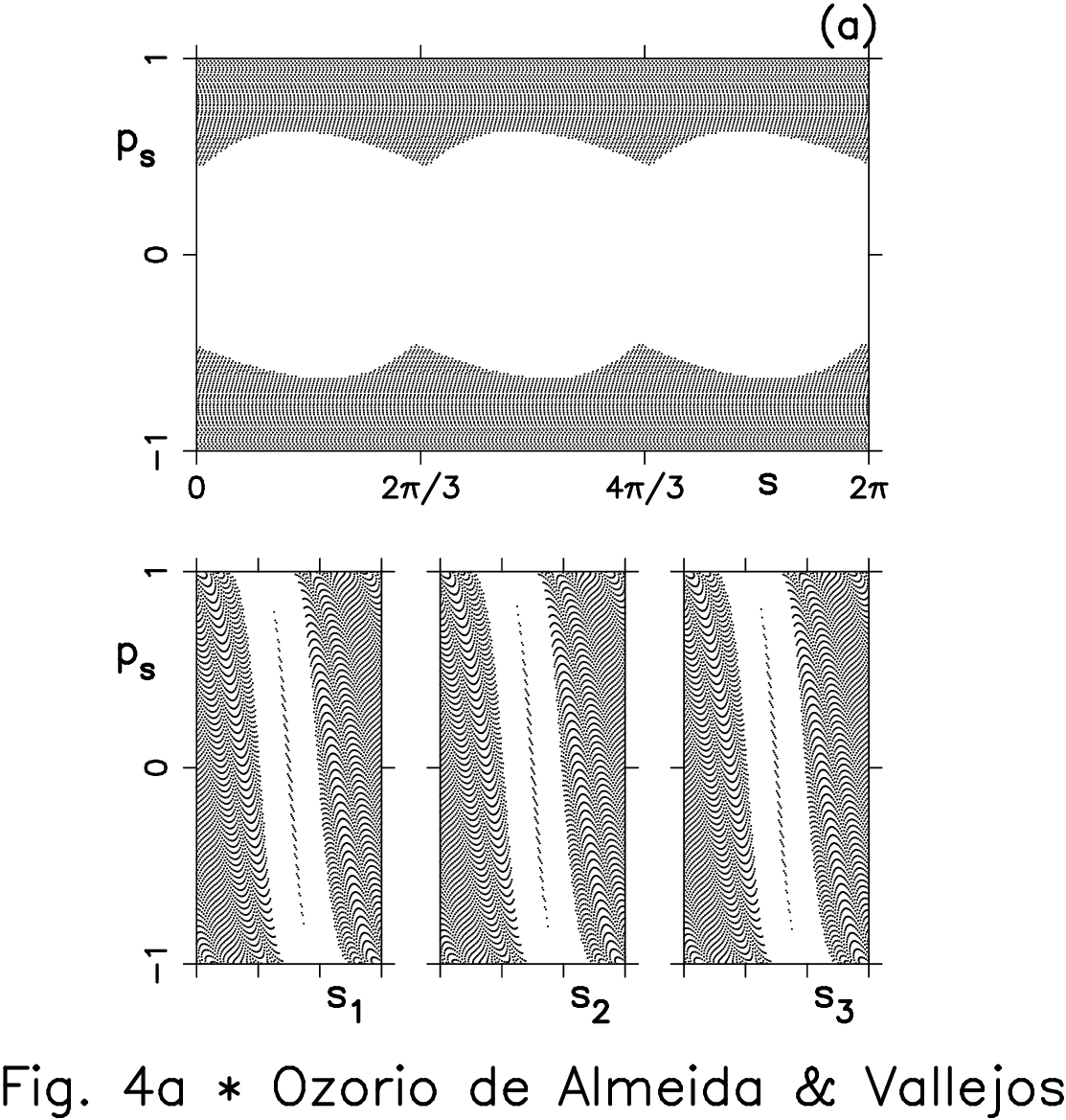}
\end{figure}

\begin{figure}[h]
\epsfxsize=8.5cm
\epsfbox[39 91 573 652]{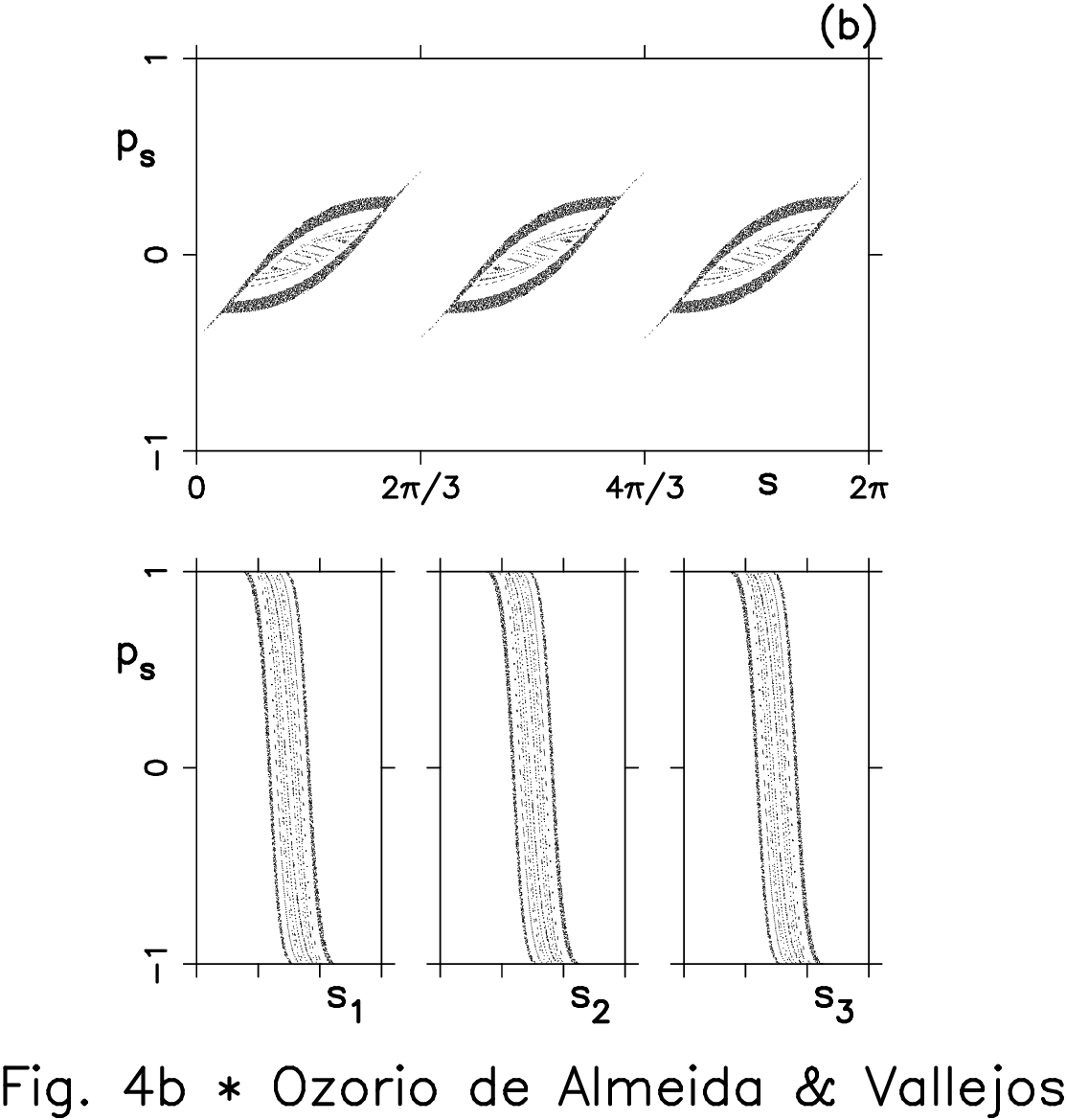}
\vspace*{1.0pc}
\caption{The return map for the three--discs system. Initial 
conditions are started at the outer circle of Fig.~3, with momenta 
pointing inwards (upper rectangles). 
(a) The first iterate of the initial rectangle is shown in dark.
Accordingly, white regions correspond to the first iterate of the phase
space of the three discs. The dark region in the phase space of the
outer circle is associated to scattering trajectories that do not hit
the inner discs (indicated as {\bf a} in Fig.~3).
(b) After three 
iterations fractal structures begin to develop. Shown are those
trajectories--stuck to the discs--that have not escaped after three 
bounces (like {\bf b} in Fig.~3), and those which are escaping after
two bounces (upper rectangle).}
\end{figure}

\end{document}